# MODELING OF PLASMA-ASSISTED CONVERSION OF LIQUID ETHANOL INTO HYDROGEN ENRICHED SYNGAS IN THE NONEQUILIBRIUM ELECTRIC DISCHARGE PLASMA-LIQUID SYSTEM

D.S. LEVKO[1], A.I. SHCHEDRIN[1], V.V. NAUMOV[1,2], V.YA. CHERNYAK[3],
V.V. YUKHYMENKO[3], I.V. PRYSIAZHNEVYCH[3], S.V. OLSZEWSKI[3]

[1]Institute of Physics, Ukrainian Academy of Sciences, Kiev, Ukraine,
ashched@iop.kiev.ua
[2]Institute of Fundamental Problems for High Technology, Kiev, Ukraine,
naumov@ifpht.kiev.ua
[3]Radiophysical Faculty, Kiev National T. Shevchenko University, Kiev, Ukraine,
chern@univ.kiev.ua

## Introduction

Today, there is a great interest in biofuels as an alternative to traditional fossil fuels and natural gas. Bio-ethanol can be a good candidate since it can be obtained in sufficient amounts from agricultural biomass. However, pure ethanol (ethyl alcohol $C_2H_5OH$) has a set of physicochemical limitations including a relatively low heat of combustion and low speed of ignition. One possible way is to use plasma-chemical conversion of ethanol into hydrogen-enriched synthesis gas (syngas) improving fuel combustibility and reducing pollution exhaust. There are various methods of plasma fuel conversion using thermal and non-thermal plasma. Thermal plasma, which is thermodynamically equilibrium, has characteristics of high ionization and higher energetic density that has merits of fast decomposition but demerits of poor selectivity of chemical transformations and high expenditure of power (> 1 kW). Non-thermal plasma, which is kinetically non-equilibrium, has characteristics of low ionization but benefits of high reactivity and selectivity of transformations, providing high enough productivity at relatively low power consumption (< 1 kW) that can be obtained by a high voltage discharging in a flow at atmospheric pressure. Applying non-thermal plasma, different electric discharge techniques were used e.g. corona, spark, MW, RF, DBD, gliding arc and various hybrids. In fact, all systems have their merits and demerits depending on the use [1]. Among possible types, one specific case is of our current research interest in Kiev [2–6]. It is a dynamic plasma-liquid system (PLS) using nonequilibrium plasma of the electric discharge in a gas channel with liquid wall (DGCLW). It can provide simultaneously a high level of nonequilibrium and high density of reacting species. Its main features are: (i) well-developed plasma-liquid interface and large surface-to-volume ratio; (ii) wide possibilities of control of plasma-created gas- and liquid-phase components during the processing; (iii) possibility to work in the cw regime using both dc and ac modes. The main idea is that the DGCLW can be burning within liquid fuel without preliminary gasification. In this work we report recent results of our experimental and theoretical studies related to plasma conversion of ethanol into syngas in the DGCLW PLS using available diagnostics and numerical modeling.

## Experiments and modeling

Experiments were done in the PLS reactor of the DGCLW type [5] using atmospheric air injected in the ethanol-water solution in the gap between the electrodes, so the discharge was burned in the gaseous cavity in the mixture of air and ethanol-water vapors. Fig. 1 shows a schematic of the DGCLW for two different schemes of electrodes: (a) solid electrode + liquid under positive (+) or negative (-) potential, and (b) two solid electrodes (coaxial cooper rods) immersed in liquid when a gas channel formed by the counter-flow air streams between the electrodes. In all cases the discharge worked in the continuous regime powered by the high voltage dc supply. Fig. 2 shows typical current-voltage characteristics of the DGCLW in the ethanol-water solution 5:1 for three different modes: (1) solid electrode + liquid cathode, (2) solid electrode + liquid anode, and (3) two solid electrodes. The dropping character of *I-V* curves at currents from 50 to 300 mA indicates the transition regime from the abnormal glow to the arc discharge. The enhanced voltage up to 3.5 kV in case of liquid electrodes is associated with the additional potential drop across a liquid and weaker secondary ion-electron emission as compared to metal electrodes.

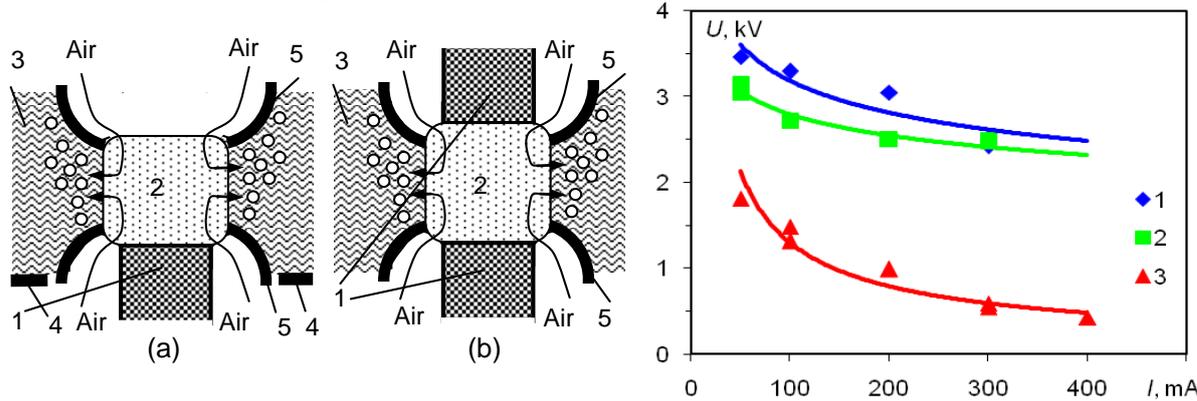

Fig. 1. Basic schemes of the DGCLW: 1 – copper rod electrodes, 2 – discharge plasma, 3 – liquid ethanol, 4 – metallic flange, 5 – quarts tubes

Fig. 2. Current-voltage curves of DGCLW: 1 – liquid cathode; 2 – liquid anode; 3 – two solid electrodes. $C_2H_5OH:H_2O = 5:1$

In experiments, the gap between electrodes, discharge currents, air flow rates, ethanol-water mixing ratios and processing time were varied to optimize the process. In nominal regimes, the discharge power did not exceed 100–300 W. The temperature of the work solution in the reactor was measured by the immersed thermocouple. It was found that the time of the temperature stabilization depended on the composition of the solution: in case of pure ethanol it needs 1–1.5 min, in case of pure water in needs 6–8 min.

The plasma conditions in the discharge during the processing were diagnosed by optical emission spectroscopy [5] using a high-speed CCD-based UV-NIR diffraction spectrometer Solar SL40-3648. The measurements have shown that the DGCLW spectra are multi-component and contains molecular bands of OH UV-system ($A^2\Sigma^+$-$X^2\Pi$: (0-0) 306.4-308.9 nm), N$_2$ 2$^+$-system ($C^3\Pi_u$-$B^3\Pi_g$: (0-0) 337.1, (0-1) 357.7, (0-2) 380.5, (1-4) 399.8 nm), CN ($B^2\Sigma^+$-$X^2\Sigma^+$: (0-0) 388.3 nm), C$_2$ Swan band ($d^3\Pi_g$-$a^3\Pi_u$: (0-0) 516.5 nm, (1-1) 512.9 nm, etc); and atomic lines of HI (656.3, 486.1 nm), OI (777.1, 844.6, 926.6 nm), and CuI (324.7, 327.4, 465.1, 510.5, 515.3, 521.8 nm) which identify a nonequilibrium nature of the DGCLW plasma. The H Balmer series emission indicates the vaporization and dissociation of ethanol and water molecules and production of hydrogen, the C$_2$ emission relates to the carbon formation, the N$_2$ and CN emissions relate to the dissociating air. The OES simulations of selected spectral lines (Cu, O, H) and bands (N$_2$, OH, CN) give the character electronic and vibrational temperatures in the range of 0.5 eV ($T_e$) and 0.35 eV ($T_v$).

The component content of the output synthesis gas products after the processing was analyzed by mass-spectrometry using a monopole mass-spectrometer MX 7301 and by gas chromatography using a gas chromatograph 6890 N Agilent [5]. The measurements have shown that gas-chromatography and mass-spectrometry data correlate well, and that $H_2$ and CO are the main components of synthesis gas produced from ethanol in the DGCLW. The fractional amount of $H_2$ and CO in the syngas reaches ~87–89 % that is many times higher than for all hydrocarbons including $CH_4$, $C_2H_2$, $C_2H_4$, and $C_2H_6$ (the total syngas amount in the output gas products is about 30 % by volume). With increasing discharge power and with water dilution, the $H_2$ yield slightly increases whereas the $O_2$ content slightly decreases, and the $N_2$ content changes non-monotonically. The maximal yield of $H_2$ is revealed if ethanol and water in the mixture are in equal amounts.

Numerical modeling of the process in the DGCLW PLS was made using the system of kinetic equations for kinetically valuable components in air-ethanol-water vapor plasma and the Boltzmann electron energy distribution function similarly to the fluid (volume averaged) model used in [6–7]. In the model statement it is supposed that (i) discharge is burning in the cavity with a radius that equals to the radius of the electrode tubes and with a length that equals to the gap between the electrodes; (ii) electric field in the discharge does not vary in space and time; (iii) after the pass of the gas through the discharge into the liquid its content in the discharge is totally refreshed and its flow rate in the reactor is the same as in the discharge. In calculations, the complete time of the discharge operation is divided into the equal time intervals which duration is determined by the cavity filling time. In the given case, this time is determined only by the time of the gas flowing that equals to the ratio of the cavity volume to the gas flow rate. This allows doing plasma-chemical calculations in the discharge during the one time interval only as the concentrations of components in the every time interval come to the same values, and the previous periods did not influence on the subsequent periods. According to the model, reactive plasma is characterized by two temperatures: the electron temperature $T_e$ is determined by the applied electric field $E/N$ and calculated EEDF; the gas temperature $T$ is determined by the temperature conditions of surrounding gas-liquid medium. The full kinetic mechanism includes 65 species ($C_2H_5OH$, $H_2O$, $N_2$, $O_2$, $H_2$, CO, $CO_2$, etc.), 76 electron-impact processes, and 364 chemical reactions (details are available at http://www.iop.kiev.ua/~plasmachemgroup).

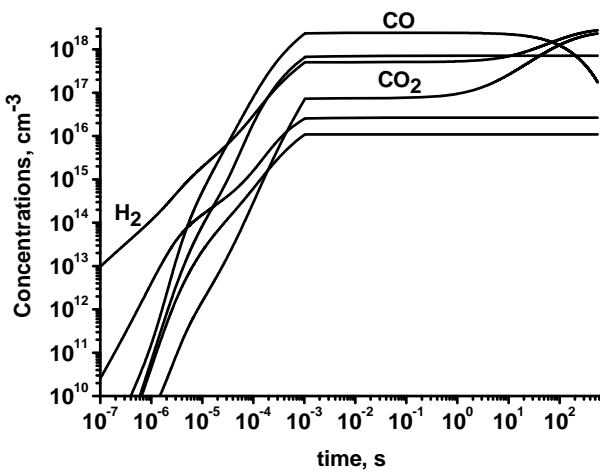 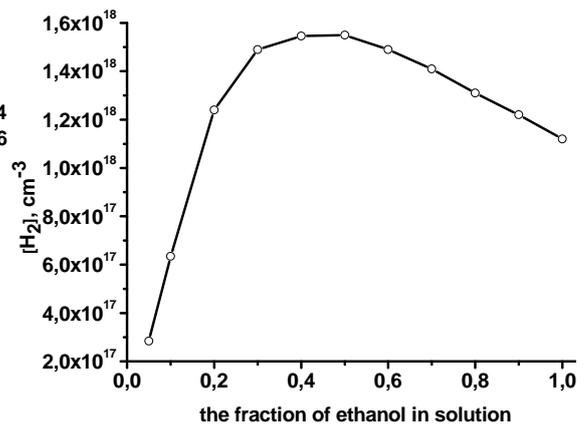

Fig. 3. Formation of synthesis gas components      Fig. 4. $H_2$ yield vs. ethanol-water ratio

Fig. 3 shows results of modeling of dynamics of formation of basic components of synthesis gas during the processing of ethanol in the DGCLW PLS (calculated for $C_2H_5OH:H_2O = 1:1$, $I = 200$ mA, $G = 55$ cm$^3$/s, $T = 323$ K). It is seen that during the discharge time up to ~$10^{-3}$ s, the concentrations of $H_2$, CO and other species steeply grow with the residence time. Outside the discharge, during the time period from $10^{-3}$ to ~10 s (the time of the gas output from the reactor) they have no dramatic changes. The final transformation of CO and $CO_2$ at the end of the process is related to the water-gas shift (WGS) reaction

$$H_2O + CO \rightarrow CO_2 + H_2 \ (\Delta H = -41 \text{ kJ/mol}). \tag{1}$$

Fig. 4 depicts the $H_2$ output depending on the ethanol-water ratio. As seen it reaches maximum at equal amounts of ethanol and water. This is explained by concurrent reactions of the hydrogen generation and e-impact dissociation of ethanol and water molecules:

$$C_2H_5OH + H \rightarrow CH_3CH_2O + H_2, \tag{2}$$

$$C_2H_5OH + e \rightarrow CH_3CH_2O + H + e, \tag{3}$$

$$H_2O + e \rightarrow OH + H + e. \tag{4}$$

Fig. 5 presents results of calculations of the component content of the synthesis gas at the outlet of the DGCLW performed for two values of medium temperature, $T = 323$ K and 355 K, in comparison with experimental data at different discharge currents $I = 200$, 300 and 400 mA. Despite the scatter of the data, one can see that in the case of $T = 323$ K (as measured by the thermocouple) the calculated concentrations of the main syngas components, $H_2$ and CO, are close to experimental values, whereas in the case of $T = 355$ K (as assumed for the boiling 50 %-ethanol-water solution) the concentrations of syngas components differ. At that, the content of $H_2$ does not vary very much, because, on one side, the water vapor content that plays a dominant role in the $H_2$ production outside the electric discharge in the volume increases leading to increasing of the $H_2$ yield; on the other side, the time of the air flowing through the volume decreases, therefore, the $H_2$ yield decreases too. Under the influence of complementary factors, the $H_2$ content keeps at the same level.

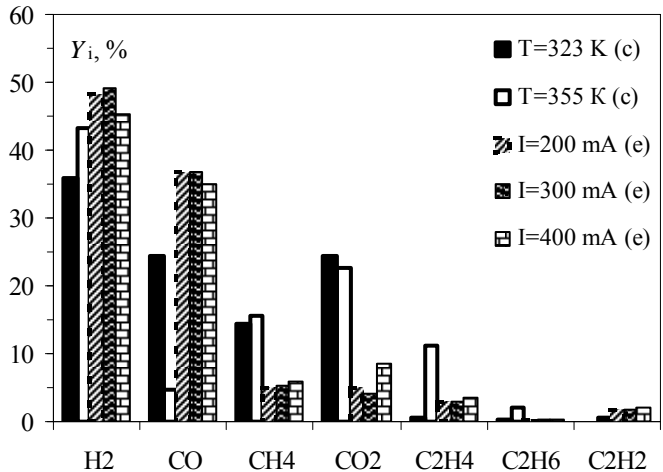 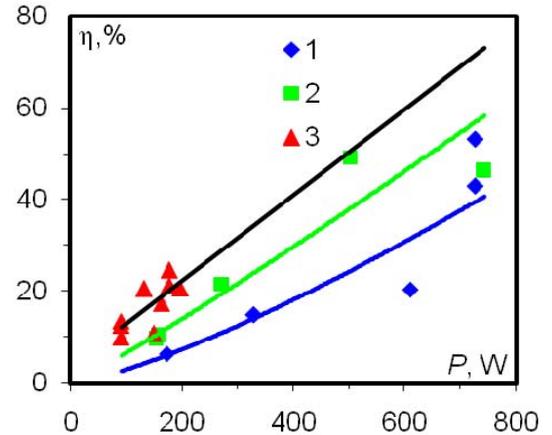

Fig. 5. Component content of synthesis gas products       Fig. 6. Efficiency of ethanol conversion

The efficiency of the conversion of ethanol into syngas by the DGCLW was estimated on the basis of thermochemical calculations using criteria: (a) energy cost of 1 m³ syngas, (b) productivity of conversion, and (c) specific heat of 1 m³ syngas combustion, utilizing the parameter of efficiency from Fulcheri [1]:

$$\eta = \frac{(Y_{H2} + Y_{CO}) \times LHV(H_2)}{IPE + Y_{HC} \times LHV(HC)} \tag{5}$$

Here, *IPE* is the input plasma energy; *Y* and *LHV* are the molar fraction and lower heating value of syngas components; *HC* means the hydrocarbon fuel (ethanol). The formula (5) assumes that CO can be totally transformed into $H_2$ by the WGS process (1) with zero energy cost. Fig. 6 demonstrates results of estimations of the conversion efficiency of the DGCLW in ethanol-water solution 5:1 for three different modes with (1) liquid cathode, (2) liquid anode, and (3) two solid electrodes. One can see that the energy efficiency of the proposed method grows with the electric power up to ~50–55%. These numbers correlate with out earlier results [4–5] and comparable with the best known plasma-aided ethanol reforming processes [1].

## Conclusions

- A dynamic PLS with the DGCLW is quite efficient in plasma conversion of ethanol into syngas. The most effective is the DGCLW with impact air jets between the electrodes.
- The main components of syngas produced from ethanol in the DGCLW are $H_2$ and CO which relative yield is many times higher than for other hydrocarbons $CH_4$, $C_2H_2$, $C_2H_4$, and $C_2H_6$. The net yield of $H_2$ increases with increasing electric discharge power and reaches maximum if ethanol and water in the solution are in equal amounts.
- The numerical plasma-chemical kinetic modeling in air-water-ethanol vapors is in a fairly good agreement with experimental data for the main components of plasma and syngas.
- The minimal value of the electric power consumption in the investigated discharge regimes is ~2 kWh/Nm$^3$ whereas the obtained syngas power is ~4 kWh/Nm$^3$.

Although there is a lot of research needed before such technology can be made technically viable, this non-thermal plasma-fuel reforming process looks very promising.

## Acknowledgements

The work was supported in part by the Ukrainian Ministry of Education and Science, by the Academy of Sciences of Ukraine and by the Science & Technology Center in Ukraine.